\documentclass[final,5p,times,twocolumn]{elsarticle}

\usepackage{graphics}

\usepackage{amssymb}
\usepackage{bm}

\journal{Physica E}

\begin{document}

\begin{frontmatter}



\title{Elementary excitations and specific heat in quantum sine-Gordon spin chain KCuGaF$_6$ }


\author[label1]{Izumi Umegaki, Toshio Ono, Hidekazu Tanaka}
\author[label2]{Masaki Oshikawa}
\author[label3]{Hiroyuki Nojiri}

\address[label1]{Department of Physics, Tokyo Institute of Technology, Oh-okayama, Meguro-ku, Tokyo 152-8551, Japan}
\address[label2]{Institute for Solid State Physics, The University of Tokyo, Kashiwanoha, Kashiwa, Chiba 277-8581, Japan}
\address[label3]{Institute for Material Research, Tohoku University, Katahira, Aoba-ku, Sendai 980-8577, Japan}

\begin{abstract}
Elementary excitations of an $S{=}1/2$ antiferromagnetic Heisenberg chain KCuGaF$_6$ were investigated through specific heat and electron spin resonance (ESR) measurements. In this compound, a staggered field is induced perpendicular to the external field because of the alternating $g$ tensor and the Dzyaloshinsky-Moriya interaction with an alternating $\bm D$ vector. Such a spin system can be mapped onto the quantum sine-Gordon (SG) model, when subjected to the external magnetic field. 
Specific heat shows clear evidence of the field-induced gap, which is related to the elementary excitations, solitons and breathers, characteristic of the quantum SG model. $q\,{=}\,0$ excitations originated from solitons and breathers were directly observed by high-frequency high-field ESR. These experimental results are well described by the quantum SG field theory.  
\end{abstract}


\end{frontmatter}


\section{Introduction}
\label{sec:intro}

The study of $S{=}1/2$ antiferromagnetic Heisenberg chain (AFHC) has a long history. The ground state energy, dispersion relation for spinon excitations and magnetization process were exactly calculated using the Bethe Ansatz \cite{Hulthen,dCP,Griffiths}. The energy of the lowest spinon excitation called des Cloizeaux-Pearson (dCP) mode is given by $E(q)\,{=}\,({\pi}/2)J{\mid}\sin q\,{\mid}$, which is a factor $\pi/2$ as large as the result of the linear spin wave theory \cite{dCP}. The dCP mode is gapless at wave vectors $q\,{=}\,0$ and $\pi$. Under magnetic field, the gapless excitations occur at incommensurate wave numbers $q\,{=}\,{\pm}\,2{\pi}m(H)\ ({\equiv}{\pm}q_0)$ and ${\pi}\,{\pm}\,q_0$ in addition to at $q\,{=}\,0$ and ${\pi}$, where $m(H)$ is the dimensionless magnetization per site \cite{Ishimura}. 

\begin{figure}[htbp]
\begin{center}
 \includegraphics[scale =0.45]{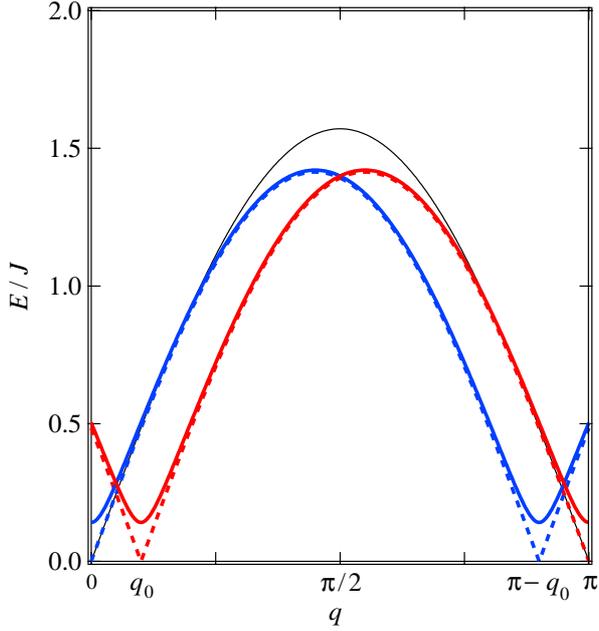}
\end{center}
\caption{Illustration of the lowest energy excitations of model (\ref{eq:model1}) under nonzero magnetic field with finite staggered field (thick solid lines). Excitations at $q\,{=}\,0, {\pi}$ and incommensurate wave numbers $q\,{=}\,{\pm}\,q_0$ and  ${\pi}\,{\pm}\,q_0$ have finite gaps. The excitations without staggered field are denoted by dashed lines. Thin solid line is the dCP mode at zero field.} 
\label{fig:excitation1}
\end{figure}

Oshikawa and Affleck \cite{Oshikawa1,AO} discussed the excitations in $S{=}1/2$ AFHC under the staggered magnetic field $h$ that is induced perpendicular to the external magnetic field $H$. The Hamiltonian of such system is expressed as
\begin{equation}
\mathcal{H}=\sum_{i} \left\{J\bm S_i\cdot \bm S_{i+1}-g{\mu}_{\rm B}HS_i^z-(-1)^ig{\mu}_{\rm B}hS_i^x\right\}. 
\label{eq:model1}
\end{equation}
In real magnetic materials, the alternating $g$ tensor and the Dzyaloshinsky-Moriya (DM) interaction with the alternating $\bm D$ vector can produce the staggered field.
Using the bosonization technique, they \cite{Oshikawa1,AO} argued that the model (\ref{eq:model1}) can be mapped onto the quantum sine-Gordon (SG) model with Lagrangian density
\begin{eqnarray}
\mathcal{L}=\frac{1}{2}\left[\left(\frac{{\partial}{\phi}}{{\partial}{t}}\right)^2\,{-}\,{(vJ)}^2\left(\frac{{\partial}{\phi}}{{\partial}{x}}\right)^2\right]+hC\cos (2{\pi}R{\tilde \phi}),
\label{eq:Lag}
\end{eqnarray}
where $\phi$ is a canonical Bose field, $\tilde \phi$ is the dual field, $R$ is the compactification radius, $v$ is the dimensionless spin velocity and $C$ is a coupling constant.
The first term corresponds to the free boson field that represents Tomonaga-Luttinger (TL) liquid. The second term expresses the nonlinear effect due to the staggered field. Oshikawa and Affleck \cite{Oshikawa1,AO} showed that all the gapless points at zero field become gapped in finite field as shown in Fig. \ref{fig:excitation1}, and that the magnitude of the gap is proportional to $H^{\,2/3}$ for $g{\mu}_{\rm B}H/J\,{\ll}\,1$. Their result gives a good description of the unexpected field-induced gap observed in Cu(C$_6$H$_5$COO)$_2$$\cdot$3H$_2$O abbreviated as Cu benzoate \cite{Dender}.

Besides Cu benzoate \cite{Dender,Asano,Nojiri}, PM$\cdot$Cu(NO$_3$)$_2\cdot$(H$_2$O)$_2$ (PM =\,pyrimidine) \cite{Feyerherm,Zvyagin} and Yb$_4$As$_3$ \cite{Oshikawa,Matysak} have been known as the quantum SG systems. In these compounds, the exchange interaction is order of 10 K and the proportional coefficient $c_{\rm s}\,{=}\,h/H$ is rather small, $c_{\rm s}\,{=}\,0.08$ \cite{Nojiri,Zvyagin}. For the deep understanding of the systems represented by the model (\ref{eq:model1}), new compounds having different interaction constants are necessary. In this paper, we introduce KCuGaF$_6$, which can be described by the model (\ref{eq:model1}) with a large exchange interaction $J/k_{\rm B}\,{\simeq}\,100$ K and a large proportional coefficient, $c_{\rm s}\,{\simeq}\,0.2$ \cite{cs}.

\begin{figure}
\begin{center}
\includegraphics[width=7.5 truecm]{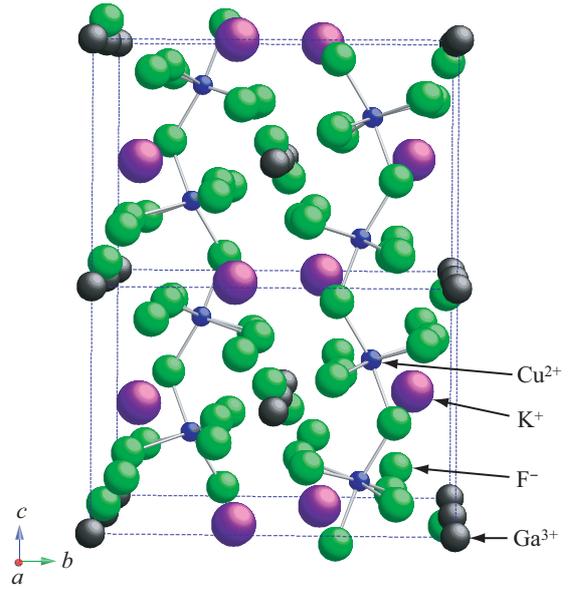}
\end{center}
\caption{Crystal structure of KCuGaF$_6$ viewed along the $a$ axis. Cu$^{2+}$ ions with spin-1/2 form chains along the $c$ axis.}
\label{fig:Crystal}
\end{figure}

KCuGaF$_6$ belongs to a pyrochlore family represented by a chemical formula AMM$^{\prime}$F$_6$, where A is a monovalent alkaline ion, and M and M$^{\prime}$ are divalent and trivalent metal ions, respectively. M$^{2+}$ and M$^{\prime}$$^{3+}$ ions form a pyrochlore lattice. According to the combination of the M and M$^{\prime}$ ions, the system shows a variety of physical properties. KCuGaF$_6$ has a monoclinic structure of space group $P2_1/c$\,\cite{Dahlke}. 
The lattice parameters at room temperature are $a\,{=}\,7.2856$\,{\AA}, $b\,{=}\,9.8951$\,{\AA}, $c\,{=}\,6.7627$\,{\AA} and $\beta\,{=}\,93.12^\circ$. Figure \ref{fig:Crystal} shows the crystal structure of KCuGaF$_6$. Cu$^{2+}$ and Ga$^{3+}$ ions are arranged to form chains along the $c$ and $a$ axes, respectively. The chains of Cu$^{2+}$ ions with spin-$1/2$ are separated by the chains of nonmagnetic Ga$^{3+}$ ions. Cu$^{2+}$ is surrounded octahedrally by six F$^-$ ions, and CuF$_6$ octahedra are elongated perpendicular to the chain direction parallel to the $c$ axis owing to the Jahn-Teller effect. The elongated axes alternate along the $c$ axis. For this reason, the hole orbitals of Cu$^{2+}$ ions are linked along the chain direction through the $p$ orbitals of F$^-$ ions. The bond angle ${\alpha}$ of the exchange pathway Cu$^{2+}$$\,{-}\,$F$^{-}$$\,{-}\,$Cu$^{2+}$ is ${\alpha}\,{=}\,129^{\circ}$. This large bond angle produces the strong antiferromagnetic exchange interaction of the order of 10$^2$ K. Thus, KCuGaF$_6$ can be expected to be $S\,{=}\,1/2$ AFHC, which can be verified from the fact that no magnetic ordering is observed down to 0.5 K \cite{Morisaki}.

In KCuGaF$_6$, the local principal axes of CuF$_6$ octahedra are tilted alternately along the $c$ axis, as shown in Fig.\ \ref{fig:Crystal}. This leads to the staggered inclination of the principal axes of the $g$ tensor. The DM interaction of the form ${\bm D}_{i}\,{\cdot}\left[{\bm S}_{i}\,{\times}\,{\bm S}_{i+1}\right]$ also exist, because there is no inversion center at the middle point of two adjacent spins along the $c$ axis. Therefore, the magnetic model of KCuGaF$_6$ in external magnetic field $\bm H$ is expressed as
\begin{eqnarray}
\mathcal{H}=\sum_{i}\left\{J{\bm S}_i\,{\cdot}\,{\bm S}_{i+1}-\mu_{\mathrm{B}}{\bm S}_i\,{{\bf g}_{i}}\,{\bm H}+{\bm D}_{i}\,{\cdot}\left[{\bm S}_{i}\,{\times}\,{\bm S}_{i+1}\right]\right\}.
\label{eq:model}
\end{eqnarray}

The staggered $g$ tensor at the $i$-th spin site is written as 
\begin{eqnarray}
{\bf g}_{i}={\bf g}_{\rm u}+(-1)^{i}{\bf g}_{\rm s},
\label{eq:g}
\end{eqnarray}
where ${\bf g}_{\rm u}$ is the uniform $g$ tensor without nondiagonal term that is common to all the spin sites and ${\bf g}_{\rm s}$ is the staggered $g$ tensor with nondiagonal terms only. The staggered $g$ tensor contributes to the staggered magnetic field ${\bm h}^{\rm s}_i\,{=}\,(-1)^i {\bf g}_{\rm s}\,{\bm H}/g'$ that is induced perpendicular to the external magnetic field $\bm H$, where $g'$ is the uniform $g$ factor for the staggered field direction. At present, details of the $g$ tensor are not clear, because no ESR signal at conventional frequencies (\,9 or 24 GHz\,) is observed at room temperature owing to large linewidth, which should be ascribed to the DM interaction discussed below. 

The ${\bm D}_i$ vector of the DM interaction is an axial vector given by the nondiagonal components of the angular momenta of adjacent magnetic ions. Since there is the $c$ glide plane at ${\pm}\,b/4$, the $ac$ plane component of the ${\bm D}_i$ vector alternates along the chain direction, but the $b$ component does not. Thus, the ${\bm D}_i$ vector is expressed as
${\bm D}_i\,{=}\,((-1)^iD_x, D_y, (-1)^iD_z)$,
where the $x$, $y$ and $z$ axes are chosen to be parallel to the $a^*\,({\perp}\,b, c)$, $b$ and $c$ axes, respectively. If the $y$ component $D_y$ is negligible, then the ${\bm D}_i$ vector is expressed as ${\bm D}_i\,{=}\,(-1)^i{\bm D}$. According to the argument by Affleck and Oshikawa\,\cite{AO}, the effective staggered field ${\bm h}_i$ acting on ${\bm S}_{i}$ is approximated as
\begin{eqnarray}
{\bm h}_i \simeq \frac{(-1)^i}{g'}\left[\frac{g}{2J}{{\bf g}_{\rm s}\,{\bm H}+{\bm H}}\,{\times}\,{\bm D}\right].
\label{eq:hst}
\end{eqnarray}
Equation\,(\ref{eq:hst}) means that the staggered field ${\bm h}_i$ is induced perpendicular to the external magnetic field $\bm H$ and its magnitude is proportional to $H$. Hence, the effective Hamiltonian of the present system can be written as eq.\,(\ref{eq:model1}). For simplification, we set $g'\,{=}\,g$ hereafter, and we rewrite $(g'/g){\bm h}_i$ as ${\bm h}_i$. 

The arrangement of this paper is as follows: In section 2, we summarize the elementary excitations in quantum SG model. The experimental procedures are presented in section 3. The results of the specific heat and ESR measurements and discussion are presented in section 4. Section 5 is devoted to the conclusion.

\section{Elementary excitations in quantum SG model}
\begin{figure}[htbp]
\begin{center}
 \includegraphics[scale =0.45]{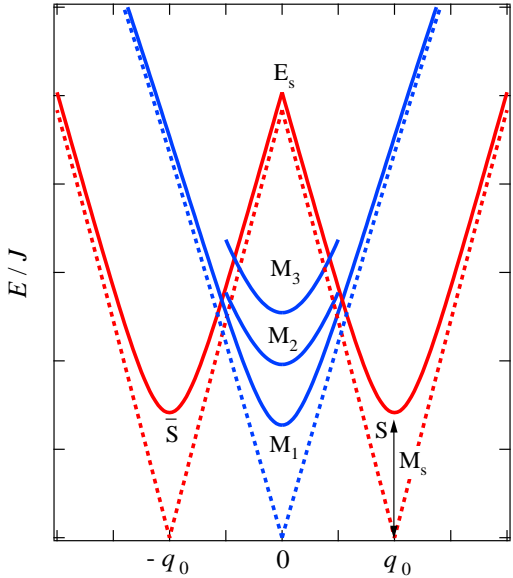}
\end{center}
\caption{Structure of low-energy excitations around $q\,{=}\,0$. Soliton, antisoliton, soliton resonance and three breathers are labeled as $S$, $\bar S$, $E_{\rm s}$ and $M_1\,{\sim}\,M_3$, respectively. The excitations without staggered field are denoted by dashed lines.} 
\label{fig:excitations}
\end{figure}

In the quantum SG model, low-energy elementary excitations are composed of solitons, antisolitons and their bound states called breathers. Figure\,\ref{fig:excitations} illustrates low-energy excitations around $q\,{=}\,0$ for $h\,{\neq}\,0$ (solid lines) and $h\,{=}\,0$ (dashed lines). Because of the staggered field $h$ induced by the external magnetic field, the gapless excitations at $q\,{=}\,0$ and ${\pm}\,q_0$ for $h\,{=}\,0$ have finite gaps. The soliton mass $M_{\rm s}$ corresponds to the excitation energy at $q\,{=}\,{\pm}\,q_0$ and ${\pi}\,{\pm}\,q_0$. 
The analytical form of $M_{\rm s}$ given by Essler {\it et al.}\,\cite{Essler3} is expressed as
\begin{eqnarray}
M_{\rm s}\,{=}\,\frac{2vJ}{\sqrt{\pi}}\frac{\Gamma \left(\displaystyle\frac{\xi}{2}\right)}{\Gamma \left(\displaystyle\frac{1+\xi}{2}\right)}\left[\frac{\Gamma \left(\displaystyle\frac{1}{1+\xi}\right)}{\Gamma \left(\displaystyle\frac{\xi}{1+\xi}\right)} \frac{c{\pi}g{\mu}_{\rm B}H}{2Jv}c_{\rm s}\right]^{(1+\xi)/2},\ \ 
\label{eq:solitonmass}
\end{eqnarray}
where $v$ is the dimensionless spin velocity, $\xi$ is a parameter given by ${\xi}\,{=}\,[2/({\pi}R^2)-1]^{-1}$ and $c$ is a parameter depending on magnetic field. The field dependences of these parameters are shown in the literature \cite{AO,Essler3,Hikihara}. For $H \rightarrow$ 0, $v\,{\rightarrow}\,{\pi}/2$, $\xi\,{\rightarrow}\,1/3$ and $c\,{\rightarrow}\,1/2$, and thus, $M_{\rm s}\,{\propto}\,H^{\,2/3}$ as shown by Oshikawa and Affleck \cite{Oshikawa1,AO}. Equation (\ref{eq:solitonmass}) is applicable in a wide magnetic field range up to the saturation field $H_{\rm s}\,{=}\,2J/g{\mu}_{\rm B}$.\\

The breathers correspond to the excitations at $q\,{=}\,0$ and ${\pi}$ and have hierarchical structures labeled by integer $n\,({=}\,1,\,2,\,\cdots)$. The mass of the $n$-th breather can be written as
\begin{eqnarray}
M_n=2M_{\rm s} {\sin}\,\left(\frac{n{\pi}{\xi}}{2}\right).
\label{eq:breather}
\end{eqnarray}
The number of breathers is limited by $n\,{\leq}\,[{\xi}^{-1}]$\,\cite{AO}. In our experimental field range, $g{\mu}_{\rm B}H/J\,{<}\,0.5$, breathers up to the third order can exist. In general, the mass of the first breather $M_1$ is smaller than the soliton mass $M_{\rm s}$, because ${\xi}\,{<}\,1/3$ in finite external field. If the SU(2) symmetry is conserved, then ${\xi}\,{=}\,1/3$ and $M_1\,{=}\,M_{\rm s}$.

In electron spin resonance experiment (ESR), we can observe only $q\,{=}\,0$ excitations. Thus, the soliton and antisoliton cannot be observed directly by ESR. Instead, we can observe a soliton resonance labeled $E_{\rm s}$ in Fig.\,\ref{fig:excitations}, which corresponds to the excitation energy at $q=0$ on the excitation branch connected to the soliton and antisoliton at $q\,{=}\,{\pm}\,q_0$\,\cite{AO,Zvyagin}.
The condition of the soliton resonance is written by 
\begin{equation}
E_{\rm s} \simeq \sqrt{M_{\rm s}^2+(g\mu_{\mathrm{B}}H)^2}.
\label{eq:soliton_resonance}
\end{equation}
From the field for the soliton resonance, we can evaluate the soliton mass $M_{\rm s}$.

Within the framework of the linear spin wave theory, we have only two modes for $q\,{=}\,0$ excitations, which are expressed as 
\begin{eqnarray}
E_-\,{\simeq}\,\sqrt{4Jc_{\rm s}HS}, \hspace{0.5cm}
E_+\,{=}\,\sqrt{{\Delta}^2+(g\mu_{\mathrm{B}}H)^2},
\end{eqnarray}
where the gap is given by ${\Delta}\,{=}\,E_-$. The modes $E_-$ and $E_+$ correspond to the first breather $M_1$ and soliton resonance $E_{\rm s}$, respectively. For the number of excitations and the field dependence of the gap, there is the significant deference between the results of the quantum SG field theory and the conventional linear spin wave theory.

\section{Experimental}
\label{sec:Exp}
KCuGaF$_6$ single crystals were grown by both vertical and horizontal Bridgman methods from the melt of an stoichiometry mixture of KF, CuF$_2$ and GaF$_3$ packed into a Pt tube. The materials were dehydrated by heating in vacuum at about 100\,$^{\circ}$C for three days. After the dehydration, one end of the Pt tube was welded and the other end was tightly folded with pliers. The temperature at the center of the furnace was set at 850\,$^\circ$C, and the lowering rate was 2$\sim$3 mm/h. KCuGaF$_6$ seems to show incongruent melting. Transparent light-pink crystals with a typical size of $3\,{\times}\,3\,{\times}\,3$ mm$^3$ were obtained. These crystals were identified as KCuGaF$_6$ by X-ray powder diffraction analysis. 

Crystallographic $a$, $b$ and $c$ axes were determined by X-ray single-crystal diffraction. Crystals are cleaved along the $(1,1,0)$ plane.
Magnetic susceptibilities measured for magnetic field parallel to these three axes are largely anisotropic below 50 K, which can be ascribed to the DM interactions \cite{Umegaki}. The magnitude of the susceptibility below 50 K is given as ${\chi}_c\,{>}\,{\chi}_b\,{>}\,{\chi}_a$. Thus, these crystallographic axes can be determined from the susceptibility measurements.

The high-frequency, high-field ESR measurements were performed in the frequency range of $135\,{-}\,761.6$ GHz using the terahertz electron spin resonance apparatus (TESRA-IMR) \cite{Nojiri} at the Institute for Material Research, Tohoku University. The temperature of the sample was lowered to 0.5 K using liquid ${}^3$He in order to suppress the finite temperature effect. Magnetic field up to 30 T was applied with a multilayer pulse magnet. FIR lasers, backward traveling wave tubes and Gunn oscillators were used as light sources. ESR absorption signals were collected for $H\,{\parallel}\,a$, $H\,{\parallel}\,b$, $H\,{\parallel}\,c$ and $H\,{\perp}\,(1,1,0)$. 
Specific heat measurements were carried out down to 0.35 K in magnetic fields of up to 9 T using a physical property measurement system (Quantum Design PPMS) by the relaxation method.

\section{Results and discussion}
\label{sec:Results}  
\subsection{Electron spin resonance measurement}

Because ESR is the most powerful tool for detecting $q\,{=}\,0$ excitations with high resolution,
we performed high-frequency ESR measurements combined with pulsed high magnetic field at 0.5 K to observe elementary excitations in KCuGaF$_6$. Due to a large exchange interaction $J/k_{\rm B}\,{=}\,103$ K, we are able to observe elementary excitations in the relatively low-field region over a wide energy range as compared with copper benzoate\,\cite{Asano,Nojiri} and PM$\cdot$Cu(NO$_3$)$_2\cdot$(H$_2$O)$_2$\,\cite{Zvyagin}. Thus, KCuGaF$_6$ is considered to be useful for comprehensive study of the elementary excitations in the quantum SG system. 
\begin{figure}[htbp]
\begin{center}
\includegraphics[width=7.8truecm,clip]{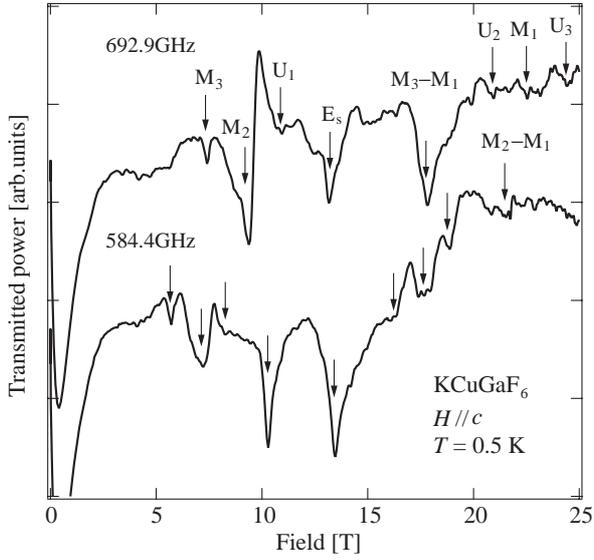}
\caption{Examples of ESR spectra obtained at 0.5 K for $H\,{\parallel}\,c$.}
\label{fig:spectra}
\end{center}
\end{figure}

Since ESR detects the $q\,{=}\,0$ excitations, the excitations labeled $M_{\rm 1}$, $M_{\rm 2}$, $M_{\rm 3}$ and $E_{\rm s}$ in Fig.\,\ref{fig:excitations} can be observed. We can also evaluate the soliton mass indirectly from the field for the soliton resonance $E_{\rm s}$. 
In Fig.\,\ref{fig:spectra}, we show examples of ESR spectra obtained at $T\,{=}\,0.5$ K for $H\,{\parallel}\,c$. Arrows indicate resonance fields in each frequency. Absorption signals observed upon sweeping field both up and down were determined as intrinsic resonance signals. In addition to the case for $H\,{\parallel}\,c$, we measured ESR spectra for $H\,{\parallel}\,b$, $H\,{\perp}\,(1,1,0)$ and $H\,{\parallel}\,a$. In these four different field directions, we observed as many as about ten resonance modes. This result is apparently different from the picture of the conventional linear spin wave theory that is composed of only two excitation modes. Labels in Fig.\,\ref{fig:spectra} denote the assignment of the modes, which will be shown below. Figure\,\ref{fig:ESR_diagram} shows the frequency vs field diagrams that summarizes the resonance data for $H\,{\parallel}\,c$ and $H\,{\parallel}\,a$. 
The resonance modes labeled as $E_{\rm s}$ and $M_n$ ($n\,{=}\,1\,{\sim}\,3$) were assigned as soliton resonance and breathers from their resonance conditions calculated using eqs.\,(\ref{eq:solitonmass})$-$(\ref{eq:soliton_resonance}) with exchange constant $J/k_{\rm B}\,{=}\,103$ K and proportionality coefficient $c_{\rm s}\,{=}\,h/H$ shown below. For $H\,{\parallel}\,c$, $H\,{\parallel}\,b$, $H\,{\perp}\,(1,1,0)$, and $H\,{\parallel}\,a$, the proportionality coefficient $c_{\rm s}\,{=}\,0.18, 0.16, 0.06$, and 0.03, respectively.
The soliton resonance and the breathers up to the third order are the main excitations predicted by the quantum SG field theory. In KCuGaF$_6$, all of these excitations were clearly observed for four different field directions. As shown in Fig.\ \ref{fig:ESR_diagram}, the experimental results are successfully described by the quantum SG field theory with only adjustable parameter $c_{\rm s}$. In these calculations, we used $g\,{=}\,2.32$ for $H\,{\perp}\,(1,1,0)$, which was determined by the present ESR measurement at $T\,{\sim}\,60$\,K. The $g$ factors used for $H\,{\parallel}\,a$, $H\,{\parallel}\,b$ and $H\,{\parallel}\,c$ are $g\,{=}\,2.28$, 2.36 and 2.12, respectively, which were determined from the uniform magnetic susceptibilities ${\chi}_{\rm u}$ at room temperature, assuming that ${\chi}_{\rm u}/g^2$ is constant.  
\begin{figure}[htbp]
\begin{center}
\includegraphics[width=8.3truecm,clip]{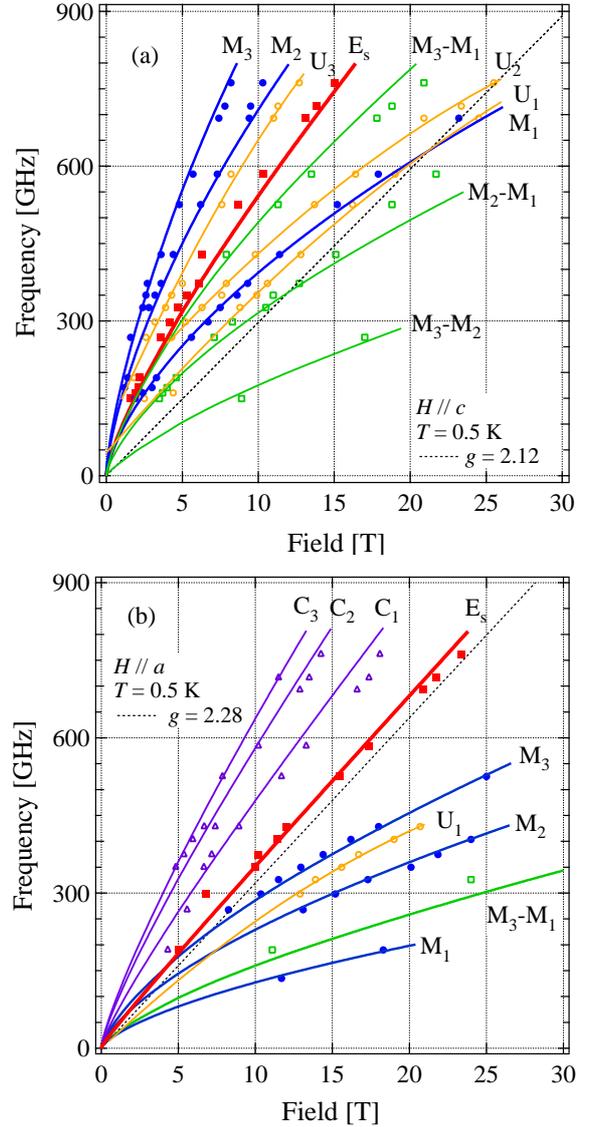}
\caption{Frequency vs field diagrams for (a) $H\,{\parallel}\,c$ and (b) $H\,{\parallel}\,a$. Symbols denote experimental results and thick solid lines are resonance conditions calculated from the quantum SG field theory with $c_{\rm s}\,{=}\,0.18$ and 0.03, respectively.}
\label{fig:ESR_diagram}
\end{center}
\end{figure}

\begin{figure}[htbp]
\begin{center}
\includegraphics[width=8.0truecm,clip]{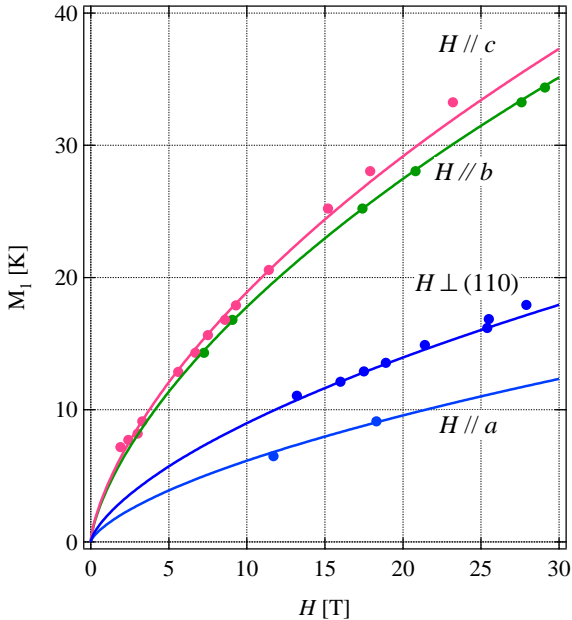}
\caption{Mass of the first brether $M_1$ obtained for $H\,{\parallel}\,c$, $H\,{\parallel}\,b$, $H\,{\perp}\,(1,1,0)$, and $H\,{\parallel}\,a$. Symbols denote experimental results and solid lines are resonance conditions calculated from eqs. (\ref{eq:solitonmass}) and (\ref{eq:breather}) with $c_{\rm s}\,{=}\,0.18, 0.16, 0.06$, and 0.03, respectively.}
\label{fig:breather}
\end{center}
\end{figure}

In Fig.\,\ref{fig:breather}, we compare the field dependence of the first breather mass $M_1$ observed for four different field directions. In the present field range $g{\mu}_{\rm B}H/J\,{<}\,0.4$, $M_1\,{\simeq}\,0.9M_{\rm s}$. The proportionality coefficient $c_{\rm s}$ reaches a maximum (0.18) for $H\,{\parallel}\,c$ and a minimum (0.03) for $H\,{\parallel}\,a$. Such a large proportionality coefficient as observed for $H\,{\parallel}\,c$ has not been observed in other SG systems. The magnitude of $c_{\rm s}$ is almost the same for $H\,{\parallel}\,c$ and $H\,{\parallel}\,b$. Since the angle between the $b$ axis and the line perpendicular to the $(1,1,0)$ plane  is ${\theta}_0\,{=}\,53.6^{\circ}$, the first breather mass $M_1$ for $H\,{\perp}\,(1,1,0)$ is approximately expressed as $M_1(H\,{\parallel}\,a)\,{+}\,{\Delta}M_1\cos^2{\theta}_0$, where ${\Delta}M_1\,{=}\,M_1(H\,{\parallel}\,b)\,{-}\,M_1(H\,{\parallel}\,a)$. This indicates that for $H\,{\parallel}\,ab$ plane, $M_1\,{\simeq}\,M_1(H\,{\parallel}\,a)+{\Delta}M_1\cos^2{\theta}$, where $\theta$ is the angle between the $b$ axis and the external field.

The intensities of the main resonance modes, the soliton resonance $E_{\rm s}$ and the breathers $M_n$, are of the same order. These two excitations occur under different conditions for the oscillating magnetic field ${H}_1$ of the submillimeter wave. The soliton resonance occurs when ${H}_1$ perpendicular to the external field $H$, while breathers are excited when ${H}_1$ is parallel to ${H}$ \cite{picture}. Since unpolarized submillimeter wave propagates in a light pipe whose diameter is larger than its wavelength, the oscillating magnetic field has components both parallel and perpendicular to the external field. Consequently, both the soliton resonance and the breathers can be observed at once in the present experiments.  

In addition to the main elementary excitations $E_{\rm s}$ and $M_n$ ($n\,{=}\,1\,{\sim}\,3$), some resonance modes were observed, as shown in Figs.\ \ref{fig:spectra} and \ref{fig:ESR_diagram}. We will discuss below these additional modes.
The excitation energies of the modes labeled $M_2\,{-}\,M_1$, $M_3\,{-}\,M_1$ and $M_3\,{-}\,M_2$ are equal to the differences between two of three breathers mass $M_n\,{-}\,M_{n'}$ calculated from eq.\ (\ref{eq:breather}). Within the framework of the quantum SG field theory, there is no excitation from the ground state that has energy $M_n\,{-}\,M_{n'}$. Thus, these resonance modes can be assigned as the interbreather transitions.
As shown in Fig.\ \ref{fig:spectra}, these interbreather transitions have sufficient intensities as the breathers. The present ESR experiments were done at 0.5 K, at which the population of the excited level is negligible under isothermal condition.
At zero magnetic field, however, breathers do not exist, there are gapless modes at $q=0$ and $\pi$, which change to the breathers modes due to the staggered field induced by the external field. The population of the gapless mode at zero magnetic field is finite even at 0.5 K. In the present ESR measurements combined with the pulsed magnetic field with the width of about 10 msec, the splitting of the levels occur under almost adiabatic conditions. Therefore, the population at zero magnetic field is maintained even in finite field and the interbreather transitions can be observed. 

As shown in Fig.\,\ref{fig:ESR_diagram} (b), weak resonance modes labeled $C_n$ with $n\,{=}\,1, 2$ and 3 were observed for $H\,{\parallel}\,a$ in the fields lower than the field for soliton resonance $E_{\rm s}$. These modes are assigned as the multiple excitations of the soliton resonance and the $n$-th breather, because their energy correspond to $E_{\rm s}\,{+}\,M_{n}$. The multiple excitation mode $C_1$ was also observed for $H\,{\perp}\,(1,1,0)$, while for $H\,{\parallel}\,b$ and $c$, no $C$ mode was observed, which should be ascribed to the large $c_s$ for the latter two field directions. For $H\,{\parallel}\,b$, we observed a resonance mode whose excitation energy is just twice as large as $M_1$ \cite{Umegaki}. This mode can be considered as the simultaneous excitation of two first breathers.  This was the first example of the two-breather resonance in the quantum SG spin system. In the present experiments, the two-breather resonance was observed only for $H\,{\parallel}\,b$. In this field direction, we observed as many as twelve modes. The energy of two-soliton excitation $2M_{\rm s}$ is almost same as the energy of the third breather $M_3$. Thus, it is hard to distinguish the $2M_{\rm s}$ mode from $M_3$ mode, although the two-soliton excitation is expected to exist.

Resonance modes $U_n$ are unknown modes, whose origins are not clear. The numbering of $U_n$ is in ascending order of excitation energy in each field direction. 
The field dependence of their energies denoted by thin solid lines in Fig.\,\ref{fig:ESR_diagram} is similar to that of mass of breathers $M_n$. In the previous measurements, we also observed the three unknown modes ($U_1\,{\sim}\,U_3$) for $H\,{\parallel}\,c$ in different specimens. Therefore, the unknown modes should be intrinsic to KCuGaF$_6$. Such unknown modes were also observed in another quantum SG system, PM$\cdot$Cu(NO$_3$)$_2\cdot$(H$_2$O)$_2$\,\cite{Zvyagin}. However, the origins of these unknown modes are unexplainable within a framework of the quantum SG field theory shown in Section 2.

\subsection{Specific heat measurement}

\begin{figure}[htbp]
\begin{center}
\includegraphics[width=8.0truecm,clip]{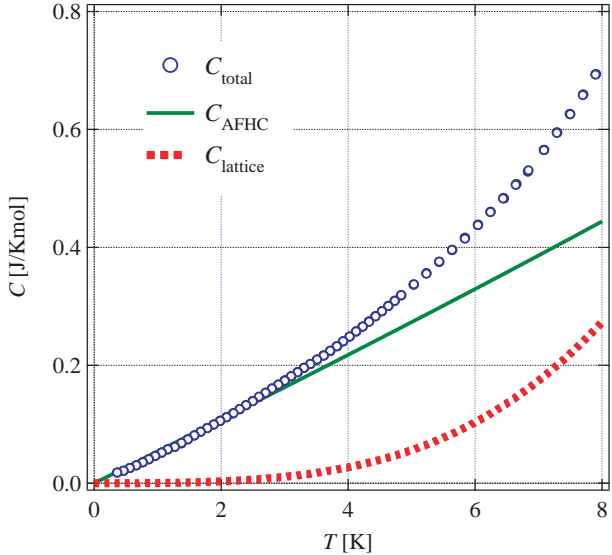}
\caption{Temperature dependence of specific heat at zero field. Open circles are total specific heat. Solid and dotted lines denote magnetic and lattice contributions, respectively.}
\label{fig:Ctotal}
\end{center}
\end{figure}

In order to study the contribution of elementary excitations to thermodynamic properties, we measured specific heat of KCuGaF$_6$. The magnetic field was applied parallel to the $c$ axis, for which the soliton gap is the largest. Figure \ref{fig:Ctotal} shows the low-temperature total specific heat $C_{\rm total}$ measured at zero magnetic field. No magnetic ordering was observed down to 0.36 K, which indicates good one-dimensionality of the present system. $C_{\rm total}$ at zero field exhibits almost linear temperature dependence below 4 K characteristic of the $S\,{=}\,1/2$ AFHC \cite{Kluemper,Johnston}. $C_{\rm total}$ is composed of magnetic $C_{\rm mag}$ and lattice $C_{\rm lattice}$ contributions. The specific heat of $S\,{=}\,1/2$ AFHC for $k_{\rm B}T/J\,{<}\,0.1$ is approximately given by \cite{Kluemper,Johnston}
\begin{eqnarray}
C_{\rm mag} = \frac{2Rk_{B}T}{3J}.
\label{eq:h_st}
\end{eqnarray}
In this low temperature region, the Tomonaga-Lutinger (TL) liquid state is realized. The exchange constant in KCuGaF$_6$ is $J/k_{\rm B}\,{=}\,103$\,K, which was obtained from the magnetic susceptibility data \cite{Morisaki,Umegaki}. Thus, the condition $k_{\rm B}T/J\,{<}\,0.1$ is satisfied for $T\,{<}\,10$\,K. The lattice contribution $C_{\rm lattice}$ shown by dashed line in Fig.\,\ref{fig:Ctotal} was obtained by subtracting the $T$-linear magnetic contribution from the total specific heat $C_{\rm total}$. The magnetic specific heat in finite magnetic field was obtained by subtracting $C_{\rm lattice}$ from the total specific heat $C_{\rm total}$.

\begin{figure}[htbp]
\begin{center}
\includegraphics[width=8.0truecm,clip]{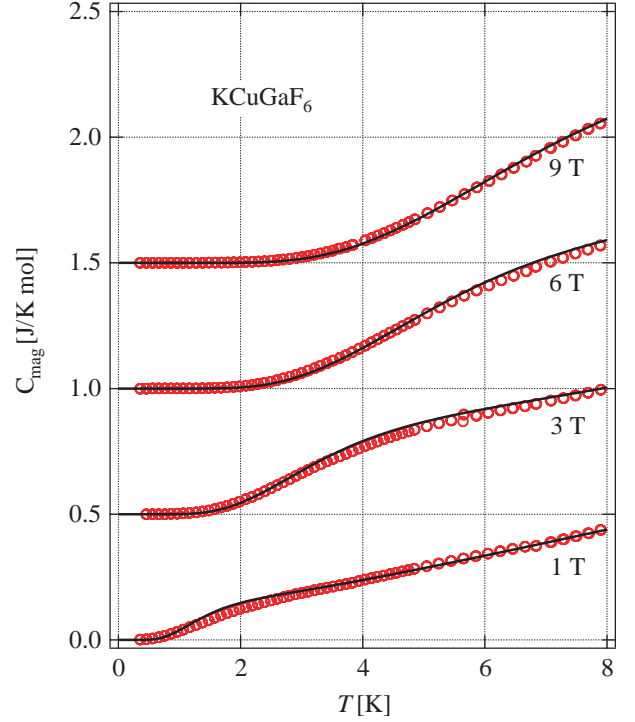}
\caption{Temperature dependence of magnetic specific heat measured at $H\,{=}\,1, 3, 6$ and 9 T for $H\,{\parallel}\,c$. Each data is shifted upward by 0.5 J/(K mol). Open circles denote experimental data and solid lines are calculations based on the quantum SG field theory with soliton mass shown in Fig.\,\ref{fig:soliton_gap}.}
\label{fig:Cmag}
\end{center}
\end{figure}

Figure\,\ref{fig:Cmag} shows the low-temperature magnetic specific heat $C_{\rm mag}$ obtained at several magnetic fields applied parallel to the $c$ axis. With increasing temperature, $C_{\rm mag}$ exhibits exponential increase, which indicates the existence of the field-induced gap. With further increasing temperature, $C_{\rm mag}$ displays a rounded shoulder and increases linearly. As magnetic field increases, the shoulder shifts to higher temperature and becomes broader. This shows that the gap increases with applied magnetic field. 

Specific heat of the quantum SG model can be obtained by solving a set of integral equations based on the Bethe Ansatz and the SU(2) symmetry \cite{Destri,Fowler}, for which the compactification radius is set as $R\,{=}\,1/\sqrt{2{\pi}}$. The calculated results were actually in agreement with experimental results in other SG model compounds, Cu benzoate \cite{Essler2} and Yb$_4$As$_3$ \cite{Oshikawa}. When the SU(2) symmetry is assumed, the mass of the first breather is the same as the soliton mass, i.e., $M_{\rm s}\,{=}\,M_1\,({\equiv}\,{\Delta})$. The solid lines in Fig.\,\ref{fig:Cmag} show the theoretical specific heat of the quantum SG model with the SU(2) symmetry \cite{Oshikawa}, see also Ref.\,\cite{Essler2}. In this calculation, the adjustable parameter is the gap $\Delta$. The behavior of the magnetic specific heat observed in KCuGaF$_6$ is well reproduced by the present analysis. With further increasing temperature above 8 K, the discrepancy between the experimental data and fitting curves becomes larger. This is because in such high temperature region, the description by the TL liquid starts to break down. 

\begin{figure}[htbp]
\begin{center}
\includegraphics[width=8.0truecm,clip]{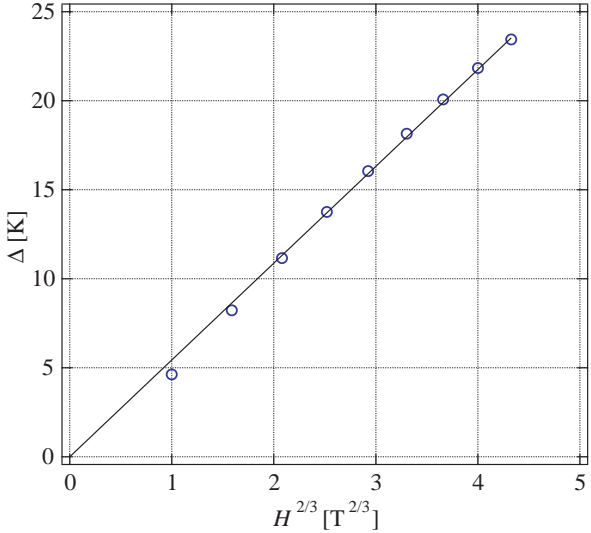}
\caption{Gap $\Delta$ as a function of $H^{\,2/3}$. Open circles denote the gaps obtained from specific heat measurements for $H\,{\parallel}\,c$. Solid line is a linear fit.}
\label{fig:soliton_gap}
\end{center}
\end{figure}

Figure\,\ref{fig:soliton_gap} shows the gap $\Delta$ as a function of $H^{\,2/3}$. It is evident that the gap is described as ${\Delta}\,{=}\,AH^{\,2/3}$. The coefficient $A$ is obtained as $A\,{=}\,5.4$\ K/(T$^{2/3}$). This field dependence of the gap is different from the result of the linear spin wave theory, which derives ${\Delta}\,{\propto}\,H^{\,1/2}$. From the field for soliton resonance in ESR measurements, we obtained soliton mass indirectly \cite{Morisaki, Umegaki}. The soliton mass calculated with eq.\,(\ref{eq:solitonmass}) and $c_{\rm s}\,{=}\,0.18$ for $H\,{\parallel}\,c$ shows the $H^{\,2/3}$ dependence with proportionality coefficient $A\,{=}\,4.3$\ K/(T$^{2/3}$). The soliton mass obtained from the specific heat result is 1.26 times as large as that obtained from the ESR measurements. The discrepancy may be ascribed to the SU(2) symmetry, i.e., $\xi \,{=}\,1/3$, which was assumed for the calculation of the specific heat. When the SU(2) symmetry is broken by the applied magnetic field, the parameter $\xi$ becomes less than 1/3. However, eq.\,(\ref{eq:solitonmass}) gives almost the same soliton mass in our experimental field range. Therefore, only from the SU(2) symmetry, we cannot explain the discrepancy between soliton masses evaluated from ESR and specific heat measurements. The presence of the unknown modes, $U_1\,{-}\,U_3$, which are not explainable in terms of the simple quantum SG model, may also be responsible for the discrepancy.
At present, we have no clear explanation about the discrepancy.   
Anyhow, in KCuGaF$_6$, the considerably large staggered field is induced when subjected in the uniform magnetic field parallel to the $c$ axis.

\section{Conclusion}
\label{sec:Conc}

In conclusion, we have presented the results of ESR measurement and the specific heat measurements on $S\,{=}\,1/2$ AFHC KCuGaF$_6$ with the large exchange interaction $J/k_{\rm B}\,{=}\,103$ K. In KCuGaF$_6$, the staggered magnetic field $h$ is induced perpendicular to the external magnetic field $H$ owing to the DM interaction with alternating $\bm D$ vectors and the staggered $g$ tensor. Thus, the present system can be represented by the quantum SG model in a magnetic field. 
In the present high-freaquency ESR measurements combined with pulsed high magnetic field, breathers up to the third order and soliton resonance concerning with soliton mass were directly observed. The energies of these elementary excitations are in good agreement with the calculations based on the quantum SG field theory with a single adjustable parameter $c_{\rm s}\,{=}\,h/H$. The proportionality coefficient $c_{\rm s}$ varies widely from 0.03 to 0.18, depending on the field direction. We also observed additional modes, i.e., inter-breather transitions, multiple excitations, and unknown modes that have no clear explanation for their origin. 

We measured specific heat for $H\,{\parallel}\,c$, where the gap is the largest. The specific heat shows the evidence of the field-induced gap, which corresponds to the soliton gap at incommensurate wave vector $q_0$ and the breather gap at $q\,{=}\,0$ and $\pi$. We analyzed the temperature dependence of specific heat, using the SG field theory based with SU(2) symmetry. We found that the gap is almost proportional to $H^{\,2/3}$, as predicted by the SG field theory. Thus, we can conclude that the quantum SG model gives a good description of the elementary excitations in KCuGaF$_6$ . Details of the present work on specific heat will be published elsewhere \cite{Umegaki2}.

\section*{Acknowledgment}
This work was supported by a Grant-in-Aid for Scientific Research (A) from the Japan Society for the Promotion of Science, and by a Global Center of Excellence Program ``Nanoscience and Quantum Physics'' at Tokyo Tech and a Grant-in-Aid for Scientific Research on Priority Areas ``High Field Spin Science in 100 T'', both funded by the Japanese Ministry of Education, Culture, Sports, Science and Technology.





\bibliographystyle{elsarticle-num}
\bibliography{<your-bib-database>}

\begin{thebibliography}{00}
\bibitem{picture} 
In the classical picture, the motion of the total magnetization corresponding to the soliton resonance is the precession around the external field $H$, and that for the breather is the oscillation parallel to $H$.
\end{thebibliography}







\end{document}